\documentclass[english,aps,prl,reprint]{revtex4-1}
\usepackage[T1]{fontenc}
\usepackage[latin9]{inputenc}
\usepackage{bm}
\usepackage{amsmath}
\usepackage{graphicx}

\makeatletter
\usepackage[all]{xy}
\usepackage{variables,new_variables,mathtools}
\makeatother

\usepackage{babel}
\begin{document}
\title{Variational Discrete Action Theory}
\author{Zhengqian Cheng and Chris A. Marianetti}
\affiliation{Department of Applied Physics and Applied Mathematics, Columbia University,
New York, NY 10027}
\date{\today}
\begin{abstract}
Here we propose the Variational Discrete Action Theory (VDAT) to study
the ground state properties of quantum many-body Hamiltonians. VDAT
is a variational theory based on the sequential product density matrix
(SPD) ansatz, characterized by an integer $\discn$, which monotonically
approaches the exact solution with increasing $\discn$. To evaluate
the SPD, we introduce a discrete action and a corresponding integer
time Green's function. We use VDAT to \textit{exactly} evaluate the
SPD in two canonical models of interacting electrons: the Anderson
impurity model (AIM) and the $d=\infty$ Hubbard model. For the latter,
we evaluate $\discn=2-4$, where $\discn=2$ recovers the Gutzwiller
approximation (GA), and we show that $\discn=3$, which exactly evaluates
the Gutzwiller-Baeriswyl wave function, provides a truly minimal yet
precise description of Mott physics with a cost similar to the GA.
VDAT is a flexible theory for studying quantum Hamiltonians, competing
both with state-of-the-art methods and simple, efficient approaches
all within a single framework.
\end{abstract}
\maketitle
Computing the ground state properties of quantum many-body Hamiltonians
is a fundamental task in physics. A common strategy to approximately
solve a Hamiltonian is the use of variational wave functions, which
allows one to find the best solution within some fraction of the Hilbert
space. A generic approach is to start from some reference wave function
and apply some projector, as in the well known Jastrow\cite{Jastrow19551479}
and Gutzwiller\cite{Gutzwiller1963159,Gutzwiller1964923,Gutzwiller19651726}
variational wave functions. A key limitation to such approaches is
that they often cannot be intelligently improved, meaning that it
is difficult to increase the searchable region of Hilbert space efficiently.
One approach to address this limitation is tensor network methods\cite{Mcculloch200710014,Schollwock2005259,Schollwock201196},
where some control parameter increases accuracy at some computational
cost; but these approaches have only proven well suited for low dimensional
systems. 

Here we propose a new class of variational density matrices: the sequential
product density matrix (SPD). The SPD is motivated by the Trotter-Suzuki\cite{Suzuki1993432}
decomposition, and is characterized by an integer $\discn$. The SPD
provides a paradigm for variational approaches in that the precision
can be systematically improved by increasing $\discn$, and it can
be applied beyond low dimensions. In practice, such an ansatz is not
useful unless one has a systematic and efficient approach for evaluating
it. Our key development is the introduction of the discrete action
theory (DAT) and the corresponding \textit{integer time} Green's function,
which may be used for evaluating an SPD. The DAT has a perfect parallel
to the standard many-body Green's function formalism, though with
non-trivial differences; such as a new discrete Dyson equation. Many
of the key ideas from traditional many-body theory can immediately
be generalized to the DAT, such as the Dynamical Mean-Field Theory
(DMFT)\cite{Georges199613}. Using the DAT for evaluating the SPD,
we can then perform the variational minimization to obtain the ground
state, and we refer to this entire approach as the Variational Discrete
Action Theory (VDAT). There is companion manuscript to this paper
which provides extensive derivations and minimal examples to document
the foundations of VDAT\cite{Cheng2020long}. 

Given a Hamiltonian $\hat{H}=\hat{H}_{0}+\hat{V}$, where $\hat{H}_{0}$
is non-interacting and $\hat{V}$ is interacting, we motivate the
SPD by considering the following variational wave function:
\begin{align}
\exp(\gamma_{1}\hat{H}_{0})\exp(g_{1}\hat{V})...\exp(\gamma_{N}\hat{H}_{0})\exp(g_{N}\hat{V})|\varphi_{0}\rangle,\label{eq:SPW}
\end{align}
where $\gamma_{i},g_{i}$ are variational parameters and $|\varphi_{0}\rangle$
is the ground state wave function of $\hat{H}_{0}$. Equation \ref{eq:SPW}
can be viewed as a variational application of the Trotter-Suzuki decomposition\cite{Trotter1959545,Suzuki1976183,Suzuki1993432},
where the $N\rightarrow\infty$ limit will cover the exact ground
state wave function. The essence of this ansatz was first proposed
several decades ago by Dzierzawa \textit{et. al} \cite{Dzierzawa19951993},
motivated by the generalization of the Baeriswyl wave function\cite{Baeriswyl19870,Baeriswyl20002033}
by Otsuka\cite{Otsuka19921645}; and all of this work was motivated
by improving upon the well known Gutzwiller wave function\cite{Gutzwiller1963159}.
More recently, a unitary version of this wave function was proposed
in the context of quantum computing by Farhi et. al\cite{Farhi1411.4028},
and further extended by Wecker \textit{et. al} \cite{Wecker2015042303}
and Grimsley \textit{et. al}\cite{Grimsley20193007}. Our SPD further
generalizes the idea behind Eq. \ref{eq:SPW}. 

Given a Hamiltonian with $L$ spin orbitals, the SPD is given as
\begin{align}
\spd=\exp\left(\sppm_{1}\cdot\spom\right)\hat{P}_{1}\dots\exp\left(\sppm_{\discn}\cdot\spom\right)\hat{P}_{\discn}=\hat{\mathcal{P}}_{1}\dots\hat{\mathcal{P}}_{\discn},
\end{align}
where $\hat{P}_{\tau}$ is a generic interacting projector, and $\tau=1,\dots,\discn$
is the integer time label; $\exp\left(\sppm_{\tau}\cdot\spom\right)$
is the noninteracting projector, where $\sppm_{\tau}\cdot\spom\equiv\sum_{i=1}^{L}\sum_{j=1}^{L}[\sppm_{\tau}]_{ij}[\spom]_{ij}$
and $[\spom]_{ij}=\hat{a}_{i}^{\dagger}\hat{a}_{j}$; and $\hat{\mathcal{P}}_{\tau}=\exp\left(\sppm_{\tau}\cdot\spom\right)\hat{P}_{\tau}$.
When using the SPD as a variational density matrix, it must be restricted
to a symmetric and semi-definite form; and there are two variants
for a given $\discn$\cite{Cheng2020long}. Referring to $\tau$ as
an integer time is likely counterintuitive, but this will prove essential
in building the DAT formalism to evaluate the SPD. It should be noted
that the most general non-interacting projector would include the
terms $\hat{a}_{i}^{\dagger}\hat{a}_{j}^{\dagger}$ and $\hat{a}_{i}\hat{a}_{j}$,
but we presently omit them for brevity. We also define a non-interacting
SPD as $\spd_{0}=\exp\left(\sppm_{1}\cdot\spom\right)\dots\exp\left(\sppm_{\discn}\cdot\spom\right)$,
which will be the starting point for the perturbative expansion of
the SPD. The variational parameters of the SPD are the $\sppm_{\tau}$
and the parameters within $\hat{P}_{\tau}$. A common choice for the
interacting projector is $\hat{P}_{\tau}=\exp(\hat{\mathcal{V}}_{\tau})=\exp\left(\sum_{i}g_{\tau,i}\hat{V}_{i}\right)$
where $\sum_{i}\hat{V}_{i}=\hat{V}$ is some decomposition of the
interacting portion of the Hamiltonian, though there are many possible
choices (e.g. as in the Jastrow wave function for the Hubbard model\cite{Jastrow19551479,fazekas19881021,Yokoyama19903669,Capello2005026406}).
The SPD brings several generalizations over Eq. \ref{eq:SPW}. First
and most obviously, the SPD explicitly includes all possible variational
freedom at the single particle level, and formally allows for a generic
interacting projector. Second, and less obviously, the SPD form allows
for a systematic evaluation using the integer time Green's function
formalism introduced in this paper. It is useful to note that $\discn=1$
recovers the well known Hartree-Fock approximation; $\discn=2$ recovers
the Gutzwiller, Baeriswyl, and Jastrow wave functions, in addition
to unitary and variational coupled cluster\cite{Bartlett198829,Bartlett1989133,Kutzelnigg1991349,Taube20063393};
and $\discn=3$ recovers the Gutzwiller-Baeriswyl\cite{Otsuka19921645}
and Baeriswyl-Gutzwiller\cite{Dzierzawa19951993} wave functions (see
\cite{Cheng2020long} for a detailed discussion). 

We now introduce the DAT to evaluate the Hamiltonian under the SPD
at a given set of variational parameters. We begin with the integer
time Green's function formalism, where the integer time evolution
in the integer time interaction representation is given as 
\begin{align}
 & \hat{O}_{I}\left(\tau\right)=\hat{U}_{I}(\tau)\hat{O}\hat{U}_{I}(\tau)^{-1},\\
 & \hat{U}_{I}(\tau)=\exp\left(\sppm_{1}\cdot\spom\right)\dots\exp\left(\sppm_{\tau}\cdot\spom\right),
\end{align}
where $\tau=1,\dots,\discn$. Taylor series expanding the interacting
projector, the expectation value of some operator $\hat{O}$ under
the SPD is given as
\begin{align}
\langle\hat{O}\rangle_{\spd} & =\frac{\sum_{n=0}^{\infty}\frac{1}{n!}\langle\textrm{T}(\sum_{\tau=1}^{\discn}\hat{\mathcal{V}}_{\tau,I}(\tau))^{n}\hat{O}_{I}(\discn)\rangle_{\spd_{0}}}{\sum_{n=0}^{\infty}\frac{1}{n!}\langle\textrm{T}(\sum_{\tau=1}^{\discn}\hat{\mathcal{V}}_{\tau,I}(\tau))^{n}\rangle_{\spd_{0}}},\label{eq:spd_expt}
\end{align}
where the quantum average is defined as $\langle\hat{O}\rangle_{\hat{\rho}}=\textrm{Tr}(\hat{\rho}\hat{O})/\textrm{Tr}(\hat{\rho})$;
the integer time ordering operator $\textrm{T}$ first sorts the operators
according to ascending integer time and then according to the position
in the original ordering of operators and finally the result is presented
from left to right; additionally, the resulting sign must be tracked
when permuting operators. It should be noted that our time convention
is opposite to the usual definition\cite{Mahan20000306463385}. Each
term in Eq. \ref{eq:spd_expt} can be evaluated via the non-interacting
integer time Green's function 
\begin{align}
 & [\boldsymbol{g}_{0}]_{k\tau,k'\tau'}=\langle\text{T}\hat{a}_{k,I}^{\dagger}\left(\tau\right)a_{k',I}\left(\tau'\right)\rangle_{\spd_{0}},\label{eq:g0_interaction}
\end{align}
using the integer time Wick's theorem\cite{Cheng2020long}. In general,
Eq. \ref{eq:spd_expt} will require the evaluation of an infinite
number of terms, but if the interacting projector is restricted to
a local subspace, we can resum the expansion into a finite number
of terms. 

For a test case, we consider the Anderson impurity model (AIM) on
a ring\cite{Cheng2020long}, which has recently been extensively studied
using density matrix renormalization group (DMRG)\cite{Barcza2019165130}.
In this case, the interacting projectors can be chosen as local to
the impurity, and the exponential can be rewritten as a sum of Hubbard
operators within the impurity as
\begin{align}
\hat{P}_{\tau} & =\exp(\mu_{\tau}\sum_{\sigma}\hat{n}_{\sigma}+u_{\tau}\hat{n}_{\uparrow}\hat{n}_{\downarrow})=\sum_{\Gamma}P_{\tau,\Gamma}\hat{X}_{\Gamma},
\end{align}
where $\hat{X}_{\Gamma}=|\Gamma\rangle\langle\Gamma|$ is a Hubbard
operator and $P_{\tau,0}=1$, $P_{\tau,\sigma}=\exp(\mu_{\tau})$,
and $P_{\tau,2}=\exp(2\mu_{\tau}+u_{\tau})$; and the subscripts $0,\sigma,2$
correspond to empty, singly occupied, and double occupied local states,
respectively. For this interacting projector, Eq. \ref{eq:spd_expt}
will have a finite number of terms and thus can be evaluated exactly. 

The computational cost of evaluating the total energy for a given
SPD in the AIM is dictated by two factors. The first cost is associated
with constructing the non-interacting integer time Green's function
for the entire system, which scales at worst as $\discn L^{3}$. Second,
there is the cost of evaluating the total energy using the integer
time Wick's theorem, which scales exponentially with the number of
integer time steps $\discn$\cite{Cheng2020long}. For the particular
cases of $\discn\le3$ and $L\approx1000$, the computational cost
is always dominated by $L$ via the first factor, and therefore in
this scenario a single evaluation of VDAT has a minimal computational
cost, given that the limitation is the diagonalization of the non-interacting
Hamiltonian. 

\begin{figure}[h]
\includegraphics[width=1\columnwidth]{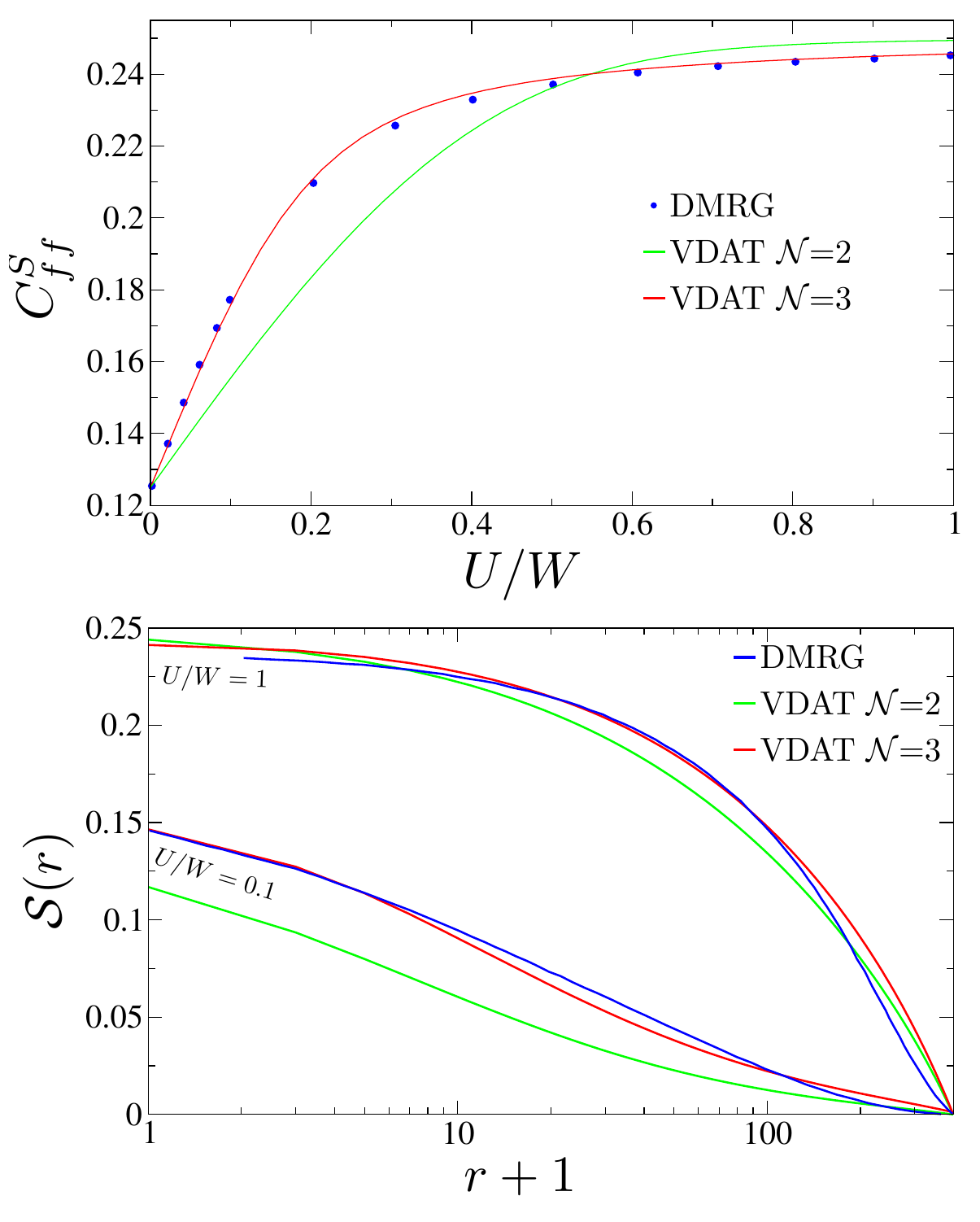}\caption{\label{fig:Cdd} A comparison of VDAT and published DMRG results\cite{Barcza2019165130}
for the Anderson impurity model on a ring with $L=1397$. (Panel a)
A plot of $C_{ff}^{S}=\langle S_{f}^{z}S_{f}^{z}\rangle=\frac{1}{4}\langle(\hat{n}_{f\uparrow}-\hat{n}_{f\downarrow})^{2}\rangle$
vs. $U/W$. (Panel b) The unscreened spin $\mathcal{S}(r)$ vs. the
distance from the impurity site.}
\end{figure}

Having all the machinery necessary to evaluate the total energy under
the SPD, we can then minimize over the variational parameters in order
to determine the ground state of the Hamiltonian. Given the specific
parameterization of the SPD we have chosen (see reference \cite{Cheng2020long},
Section VII B 1), there will be $\lceil(\discn-1)/2\rceil$ interacting
variational parameters, while there will be $3\lfloor(\discn-1)/2\rfloor$
non-interacting variational parameters. Therefore, the total number
of iterations for minimizing the energy under the SPD will be a constant
that is independent of $L$, and VDAT for $1<\discn\le3$ should have
a similar cost to $\discn=1$ (i.e. Hartree-Fock).

We now evaluate the local spin correlation $C_{ff}^{S}=\langle S_{f}^{z}S_{f}^{z}\rangle=\frac{1}{4}\langle(\hat{n}_{f\uparrow}-\hat{n}_{f\downarrow})^{2}\rangle$,
which probes the double occupancy on the impurity site (see Figure
\ref{fig:Cdd}, panel $a$), and compare to numerically exact DMRG
calculations\cite{Barcza2019165130}. For $\discn=2$, which recovers
the Gutzwiller wave function, the result is only a rough approximation
to the DMRG results. Alternatively, $\discn=3$ only shows very small
deviations from the DMRG results. Clearly, the SPD converges extremely
rapidly with respect to $\discn$. What is especially remarkable is
that $\discn=3$ has a computational cost similar to Hartree-Fock,
yet has a precision approaching DMRG. We can also compute the unscreened
spin (see Figure \ref{fig:Cdd}, panel $b$), which is a far more
challenging observable given that it involves a long range correlation
between the impurity and the bath\cite{Barcza2019165130}. Once again,
$\discn=2$ has reasonable but relatively inaccurate results, while
$\discn=3$ is quantitatively accurate.

The preceding approach of using the integer time Wick's theorem to
sum all diagrams would be intractable for a general interacting system.
This motivates us to push forward our discrete action theory, and
generalize the traditional tools of many-body physics. Consider the
interacting integer time Green's function under an SPD defined as
\begin{align}
[\boldsymbol{g}]_{k\tau,k'\tau'} & =\langle\text{T}\hat{a}_{k}^{\dagger}\left(\tau\right)\hat{a}_{k'}\left(\tau'\right)\rangle_{\spd},
\end{align}
where $\tau=1,\dots,\discn$ and $k=1,\dots,L$ and $\hat{O}(\tau)$
is an operator in the integer time Heisenberg representation defined
as 
\begin{align}
\hat{O}(\tau) & =\hat{U}_{\tau}\hat{O}\hat{U}_{\tau}^{-1}, & \hat{U}_{\tau}=\hat{\mathcal{P}}_{1}\dots\hat{\mathcal{P}}_{\tau}.\label{eq:heisenberg_rep_def}
\end{align}
Furthermore, when constructing interaction energies and computing
the gradient of the total energy with respect to the variational parameters,
we will need general integer time correlation functions $\langle\textrm{T}\hat{O}_{1}(\tau_{1})...\hat{O}_{M}(\tau_{M})\rangle_{\spd}$;
which can be rewritten in the integer time Schrodinger representation
as
\begin{alignat}{1}
 & \frac{\langle\textrm{T}\hat{\mathcal{A}}\hat{O}_{1,S}(\tau_{1})...\hat{O}_{M,S}(\tau_{M})\rangle_{\hat{1}}}{\langle\textrm{T}\hat{\mathcal{A}}\rangle_{\hat{1}}},\hspace{1em}\textrm{where}\hspace{1em}\hat{\mathcal{A}}=\hat{\mathcal{A}}_{0}\hat{P},\\
 & \hat{\mathcal{A}}_{0}=\exp\big(\sum_{\tau=1}^{\discn}\sppm_{\tau}\cdot\spom_{S}(\tau)\big),\hspace{1em}\hat{P}=\prod_{\tau=1}^{\discn}\hat{P}_{\tau,S}(\tau),\label{eq:proj_diag}
\end{alignat}
and $\hat{O}_{S}(\tau)$ is an operator in the integer time Schrodinger
representation where $\hat{O}_{S}(\tau)=\hat{O}$ after applying the
time ordering operator. We refer to $\hat{\mathcal{A}}$ as the \textit{discrete
action}, given that it encodes all possible integer time correlations
under the SPD. Moreover, we can generalize the form of $\hat{\mathcal{A}}$
such that it can describe integer time correlations beyond the SPD,
and an important generalization allows for an $\hat{\mathcal{A}}_{0}$
which has off-diagonal integer time components as 
\begin{align}
 & \hat{\mathcal{A}}_{0}=\exp(\sum_{k\tau k'\tau'}[\boldsymbol{v}]_{k\tau,k'\tau'}\hat{a}_{k,S}^{\dagger}(\tau)\hat{a}_{k',S}(\tau')),\label{eq:generalA}
\end{align}
where $\boldsymbol{v}$ is a general matrix of dimension $L\discn\times L\discn$\cite{Cheng2020long}.
We refer to this more general form as a canonical discrete action
(CDA), which will be critical to exactly evaluating the SPD in $d=\infty$.

It is useful to define the discrete generating function, which encodes
all information of the discrete action into a scalar function
\begin{equation}
Z(\boldsymbol{g}_{0})\equiv\langle\hat{\mathcal{A}}\rangle_{\hat{1}}/\langle\hat{\mathcal{A}}_{0}\rangle_{\hat{1}}.
\end{equation}
For example, using the discrete generating function and the Lie group
properties of the non-interacting many-body density matrix, we can
derive the discrete Dyson equation
\begin{align}
(\bm{g}^{-1}-\boldsymbol{1})=(\bm{g}_{0}^{-1}-\boldsymbol{1})\exp(-\boldsymbol{\Sigma})^{T},
\end{align}
where the integer time self-energy $\bm{\Sigma}$ and exponential
integer time self-energy $\boldsymbol{S}$ are obtained from the generating
function as
\begin{equation}
\exp\left(\bm{\Sigma}\right)^{T}=\boldsymbol{S}^{-1}=\boldsymbol{1}+\frac{\boldsymbol{1}}{Z\boldsymbol{1}-\frac{\partial Z}{\partial\boldsymbol{g}_{0}^{T}}\boldsymbol{g}_{0}}\frac{\partial Z}{\partial\boldsymbol{g}_{0}^{T}}.\label{eq:selfenergyviaZ}
\end{equation}
This discrete Dyson equation plays a central role in our formalism,
much like the traditional Dyson equation. In the limit of large $\discn$,
the discrete Dyson equation reverts to the usual Dyson equation assuming
that the SPD is chosen as the Trotter-Suzuki decomposition\cite{Cheng2020long}.
While we have illustrated the single particle integer time Green's
function above, any n-particle integer time correlation function can
be determined from the generating function\cite{Cheng2020long}.

\begin{figure}[h]
\includegraphics[width=1\columnwidth]{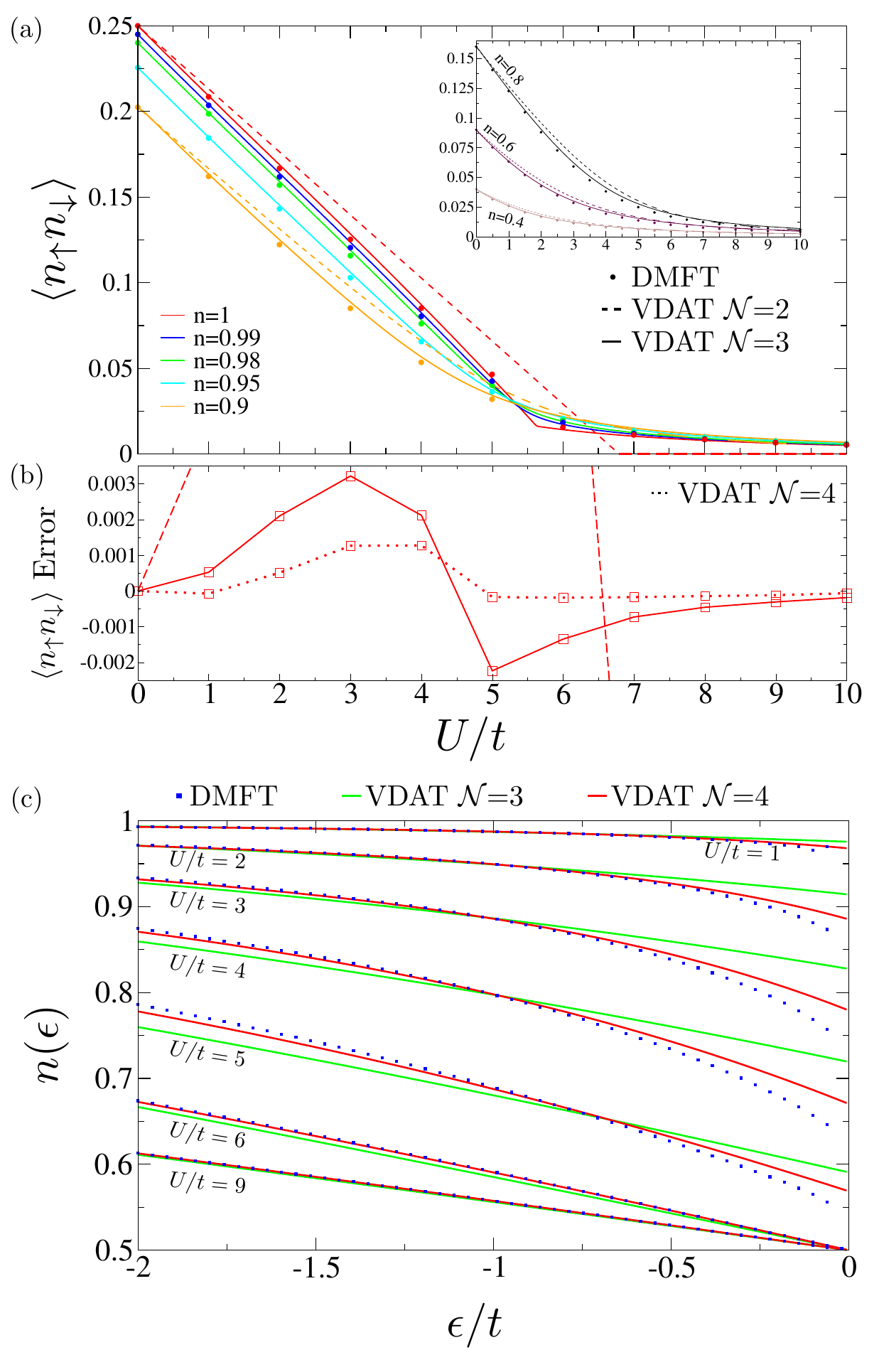}

\caption{\label{fig:docc_vs_u} Results for the $d=\infty$ Hubbard model on
the Bethe lattice. (Panel $a$) The double occupancy vs. $U/W$ for
VDAT($\discn=3$) (lines) and DMFT (points). Densities of 1, 0.99,
0.98, 0.95, and 0.9 are evaluated. (Inset) Densities of 0.8, 0.6,
and 0.4 are evaluated. (Panel $b$) The difference in the double occupancy
between VDAT($\discn=2-4$) and DMFT vs. $U/W$. (Panel $c$) The
density distribution vs. energy for various values of $U/t$ evaluated
using VDAT($\discn=3,4$) and DMFT.}
\end{figure}

Having completed the discrete action theory formalism, we now have
the proper tools to study the single-band Hubbard model, and we use
an SPD with an interacting projector $\hat{P}_{\tau}=\exp(\mu_{\tau}\sum_{i\sigma}\hat{n}_{i\sigma}+\sum_{i}u_{\tau}\hat{n}_{i\uparrow}\hat{n}_{i\downarrow})$,
while the non-interacting projector uses uses a diagonal $\boldsymbol{\gamma}_{\tau}$
in the basis that diagonalizes the non-interacting Hamiltonian. In
order to evaluate the discrete generating function, we introduce the
self-consistent canonical discrete action approximation (SCDA), which
is the integer time analogue of DMFT\cite{Georges199613}. The SCDA
maps the SPD to a collection of CDA's, with one CDA corresponding
to each site in the lattice, and the non-interacting part of the CDA
is determined self-consistently while the interacting part is taken
from the SPD. The essence of the SCDA is the assumption that the integer
time self-energy is local

\begin{align}
\bm{\Sigma}_{ij}\left(\bm{g}\right)=\delta_{ij}\bm{\Sigma}_{loc}\left(\bm{g}_{loc}\right).
\end{align}
Analogous to DMFT, which assumes a local self-energy and becomes exact
in $d=\infty$, the SCDA exactly evaluates the SPD in $d=\infty$.
For example, the SCDA for $\discn=2$ recovers the result that the
Gutzwiller approximation exactly evaluates the Gutzwiller wave function
in $d=\infty$\cite{Metzner19887382}. Additionally, the SCDA for
$\discn=3$ exactly evaluates generalizations of the Gutzwiller-Baeriswyl
and Baeriswyl-Gutzwiller wave functions in $d=\infty$, which had
not yet been achieved. For $\discn>3$, the SCDA exactly evaluates
an infinite number of variational wave functions in $d=\infty$ that
have not yet been considered. 

The SCDA algorithm exactly parallels the DMFT algorithm. We can begin
with a guess for the non-interacting integer time Green's function
$\bm{\mathcal{G}}=\int d\epsilon D\left(\epsilon\right)\bm{g}_{0}\left(\epsilon\right)$
for the CDA. We can then compute the generating function of the CDA,
which yields $\bm{S}_{loc}$. We then use this exponential integer
time self-energy to update the interacting integer time Green's function
for each energy orbital as 
\begin{align}
\bm{g}\left(\epsilon\right)=\frac{\boldsymbol{1}}{\boldsymbol{g}_{0}\left(\epsilon\right)+\left(\boldsymbol{1}-\bm{g}_{0}\left(\epsilon\right)\right)\bm{S}_{loc}}\bm{g}_{0}\left(\epsilon\right).
\end{align}
Then we obtain the new interacting local integer time Green's function
as $\bm{g}_{loc}=\int d\epsilon D\left(\epsilon\right)\bm{g}\left(\epsilon\right)$.
Finally, the new non-interacting integer time Green's function of
the CDA is 
\begin{align}
\bm{\mathcal{G}}=\bm{S}_{loc}\frac{\boldsymbol{1}}{\boldsymbol{1}+\bm{g}_{loc}\left(\bm{S}_{loc}-\boldsymbol{1}\right)}\bm{g}{}_{loc}.
\end{align}
This procedure must be iterated until self-consistency is achieved,
which yields a single evaluation of the SPD for a given set of variational
parameters. The above procedure is applicable to any Hubbard-like
model, but it will yield an \textit{exact} evaluation of the SPD for
infinite dimensions\cite{Cheng2020long}. 

The number of variational parameters for the interacting projector
will be the same as the AIM, while for the non-interacting projector
we restrict to five variational parameters for each integer time\cite{Cheng2020long}.
The main computational complexity of solving the Hubbard model as
compared to the AIM is that we must perform a self-consistency condition,
though this can normally be achieved in small number of iterations.
We now address the $d=\infty$ Hubbard model on the Bethe lattice.
It should be emphasized that the computed ground state energy within
VDAT is a rigorous upper bound for the exact ground state energy,
and we can compare to the numerically exact dynamical mean-field theory
results obtained using the numerical renormalization group (NRG) method
as the impurity solver\cite{Zitko2009085106}. We examine the double
occupancy as a function of $U/t$, where we present VDAT results for
$\discn=3$ and selected results for $\discn=2$, where the latter
is the well known Gutzwiller approximation (see Figure \ref{fig:docc_vs_u},
panel $a$). For half-filling, shown in red, we see that VDAT $\discn=3$
is very close to the DMFT solutions (points), reliably capturing the
Mott metal-insulator transition, and illustrating drastic improvement
beyond $\discn=2$. Furthermore, we can see that VDAT clearly captures
the sensitive changes with small doping, illustrated for the densities
of 1, 0.99, 0.98, 0.95, and 0.9. We can also proceed to much larger
dopings (see inset), where VDAT once again reliably describes the
DMFT solution. All VDAT results discussed thus far have been for $\discn\le3$,
and it is interesting to consider $\discn=4$ to better understand
the convergence of the VDAT. Therefore, we examine the error in the
double occupancy at half filling for $\discn=2-4$ (see Figure \ref{fig:docc_vs_u},
panel $b$). We see that $\discn=4$ has a smaller error for all values
of $U/t$, as it must, and the error for large $U/t$ is nearly zero.
Another interesting quantity is the density distribution (see Figure
\ref{fig:docc_vs_u}, panel $c$), which is more challenging given
that only the average enters the total energy. It is well know that
$\discn=2$ (i.e. Gutzwiller) produces a constant density distribution,
and we show that $\discn>2$ produces a non-trivial energy dependence.
This quantity converges less rapidly than the double occupancy, presumably
because it does not directly enter the Hamiltonian.

In conclusion, we have proven that VDAT with $\discn=3$ already yields
efficient and precise results for the Anderson Impurity model and
for the $d=\infty$ Hubbard model. Furthermore, it is straightforward
to address the multi-orbital Hubbard model\cite{Cheng2020long}, which
is under way. Given that VDAT recovers the Hartree-Fock and Gutzwiller
wave functions, it is clear that VDAT can be combined with DFT in
the same spirit as DFT+U\cite{Anisimov1997767} and DFT+Gutzwiller\cite{Deng2009075114};
and DFT+VDAT would be a prime candidate as an efficient first-principles
approach to strongly correlated materials, which should rival DFT+DMFT\cite{Kotliar2006865}.

Additionally, there are many possible directions for future development,
given the obvious parallels between the discrete action theory and
traditional many-body Green's function theory. Both diagrammatic and
auxiliary field quantum Monte-Carlo could be generalized to our formalism.
While our present work on the SPD has used real variational parameters,
we can apply VDAT using an SPD with unitary projectors\cite{Cheng2020long},
which could have utility in quantum computing\cite{Farhi1411.4028,Wecker2015042303,Grimsley20193007}
and unitary coupled cluster theory\cite{Bartlett1989133,Taube20063393,Kutzelnigg1991349}.
Finally, VDAT will be a key tool for parameterizing energy functionals
in the context of the off-shell effective energy theory\cite{Cheng2020081105},
which can be viewed as a formally exact construction based on an $\discn=2$
SPD. 

This work was supported by the grant DE-SC0016507 funded by the U.S.
Department of Energy, Office of Science. This research used resources
of the National Energy Research Scientific Computing Center, a DOE
Office of Science User Facility supported by the Office of Science
of the U.S. Department of Energy under Contract No. DE-AC02-05CH11231. 

%

\end{document}